\documentclass[conference]{IEEEtran}
\usepackage{cite}
\usepackage{amsmath,amssymb,amsfonts}
\usepackage{graphicx}
\usepackage{changes}
\definecolor{mygreen}{rgb}{0.0, 0.5, 0.0}  % Define a green color
\setaddedmarkup{\textcolor{mygreen}{#1}}
\definecolor{myred}{rgb}{0.8, 0.2, 0.2}
\setdeletedmarkup{\textcolor{myred}{\sout{#1}}}
\usepackage{textcomp}
\def\BibTeX{{\rm B\kern-.05em{\sc i\kern-.025em b}\kern-.08em
    T\kern-.1667em\lower.7ex\hbox{E}\kern-.125emX}}

\IEEEoverridecommandlockouts
\usepackage{cite}
\usepackage{tabularx}
\usepackage{mathrsfs, amsmath,amssymb,amsfonts}
\usepackage{graphicx}
\usepackage{textcomp}
\PassOptionsToPackage{table,dvipsnames,svgnames,x11names}{xcolor}
\usepackage{xcolor}
\usepackage{colortbl}

\def\BibTeX{{\rm B\kern-.05em{\sc i\kern-.025em b}\kern-.08em
    T\kern-.1667em\lower.7ex\hbox{E}\kern-.125emX}}

\usepackage{lipsum}

\usepackage[utf8]{inputenc}
\usepackage[T1]{fontenc}

\usepackage[abbreviations]{foreign}
\usepackage{algorithm}
\usepackage{algpseudocode}

\usepackage{xspace}
\usepackage{siunitx}
\usepackage{listings}
\usepackage{multirow}

\usepackage{subcaption}
\usepackage{textcomp}
\usepackage{booktabs}

\usepackage{url}
\usepackage{pifont}% http://ctan.org/pkg/pifont
\usepackage[colorlinks=true,linkcolor=blue,anchorcolor=blue,citecolor=blue,filecolor=black,menucolor=black,runcolor=black,urlcolor=blue]{hyperref}
\usepackage{orcidlink}

\usepackage[inline]{enumitem}
\setlist{noitemsep,topsep=0pt,parsep=0pt,partopsep=0pt}
\usepackage{textgreek}
\usepackage[capitalise]{cleveref}

\usepackage{tikz}
\usetikzlibrary{positioning, arrows.meta}
\usetikzlibrary{matrix}
\usepackage{pgfplots}

\definecolor{lightcolor}{rgb}{0,0.5,1}

\usepackage{fontawesome5}
\newboolean{showcomments}
\setboolean{showcomments}{true}
\ifthenelse{\boolean{showcomments}}
{ \newcommand{\mynote}[3]{
		\fbox{\bfseries\sffamily\scriptsize#1}
		{\small$\blacktriangleright$\textsf{\emph{\color{#3}{#2}}}$\blacktriangleleft$}}}
{ \newcommand{\mynote}[3]{}}

\definecolor{darkgreen}{rgb}{0.3,0.5,0.3}
\definecolor{darkblue}{rgb}{0.3,0.3,0.5}
\definecolor{darkred}{rgb}{0.5,0.3,0.3}

\newcounter{numobserv} 
\definecolor{beaublue}{rgb}{0.88, 0.93, 0.93}
\usepackage{tcolorbox}
\tcbuselibrary{skins, breakable}

\usepackage{etoolbox}
\colorlet{shadecolor}{beaublue}
\tcbset{
	colback=beaublue,
	boxrule=0.5pt,
	boxsep=0pt,
	left=1pt,
	right=1pt,
	top=1pt,
	bottom=1pt,
	sharp corners,
	colframe=white,
}

\usepackage{flushend}

\makeatletter
\let\origsection\section
\renewcommand\section{\@ifstar{\starsection}{\nostarsection}}
\newcommand\nostarsection[1]{\sectionprelude\origsection{#1}\sectionpostlude}
\newcommand\starsection[1]{\sectionprelude\origsection*{#1}\sectionpostlude}
\newcommand\sectionprelude{\vspace{0pt}}
\newcommand\sectionpostlude{\vspace{0pt}}
\let\origsubsection\subsection
\renewcommand\subsection{\@ifstar{\starsubsection}{\nostarsubsection}}
\newcommand\nostarsubsection[1]{\subsectionprelude\origsubsection{#1}\subsectionpostlude}
\newcommand\starsubsection[1]{\subsectionprelude\origsubsection*{#1}\subsectionpostlude}
\newcommand\subsectionprelude{\vspace{-2pt}}
\newcommand\subsectionpostlude{\vspace{-2pt}}
\g@addto@macro\normalsize{%
  \setlength\abovedisplayskip{2pt}
  \setlength\belowdisplayskip{2pt}
  \setlength\abovedisplayshortskip{2pt}
  \setlength\belowdisplayshortskip{2pt}
  \setlength{\floatsep}{3pt}
  \setlength{\textfloatsep}{3pt}
  \setlength{\intextsep}{2pt}
  \setlength{\dblfloatsep}{3pt}
  \setlength{\dbltextfloatsep}{3pt}
}
\makeatother
\usepackage[export]{adjustbox}
\usepackage{eso-pic}
\newcommand{\sys}{\textsc{ScamDetect}\xspace}
\newcommand{\ph}{\textsc{PhishingHook}\xspace}
\newcommand{\sysacronym}{\textbf{S}mart \textbf{C}ontract  \textbf{A}gnostic  \textbf{M}alware  \textbf{Detect}or\xspace}

\begin{document}

\title{ScamDetect: Towards a Robust, Agnostic Framework to Uncover Threats in Smart Contracts}

\author{
    \IEEEauthorblockN{
        Pasquale De Rosa~\orcidlink{0000-0001-9726-7075}\IEEEauthorrefmark{2},
        Pascal Felber~\orcidlink{0000-0003-1574-6721}\IEEEauthorrefmark{2},        
        and Valerio Schiavoni~\orcidlink{0000-0003-1493-6603}\IEEEauthorrefmark{2}        
    }
    \IEEEauthorblockA{\IEEEauthorrefmark{2}University of Neuch\^atel, Neuch\^atel, Switzerland, \url{first.last@unine.ch}}
}

\maketitle

%!TEX root = main.tex

\def\confname{55th Annual IEEE/IFIP International Conference on Dependable Systems and Networks – Supplemental Volume (DSN-S '25)}
\def\confyear{2025}
\def\confdoi{XXX}

% Copyright text: https://www.ieee.org/publications/rights/rights-policies.html
% and in the Springer's contributor consent PDF.
\definecolor{yellowPaper}{HTML}{fff8ae}
\AddToShipoutPictureFG*{%
  \AtTextUpperLeft{%
    \adjustbox{raise=3pt}{
    \begin{tcolorbox}[width=1\textwidth,colback=yellowPaper,enhanced,frame hidden,sharp corners]  
        \centering\scriptsize
        \copyright~\confyear\ 	
         by the Institute of Electrical and Electronics Engineers (IEEE). Personal use of this material is permitted. Permission from IEEE must be obtained for all other uses, in any current or future media, including reprinting/republishing this material for advertising or promotional purposes, creating new collective works, for resale or redistribution to servers or lists, or
         reuse of any copyrighted component of this work in other works.
        This is the author's version of the work.
        The final authenticated version is available online at \href{https://doi.org/10.1109/DSN-S65789.2025.00068}{https://doi.org/10.1109/DSN-S65789.2025.00068} % will be available online
        and has been published in the proceedings of the 
        \confname.
     \end{tcolorbox}} 
  }%
}%

\hypersetup{
    pdftitle={\copyright~\confyear\  Copyright 2025 by the Association for Computing Machinery, Inc. (ACM). Permission to make digital or hard copies of portions of this work for personal or classroom use is granted without fee provided that the copies are not made or distributed for profit or commercial advantage and that copies bear this notice and the full citation on the first page in print or the first screen in digital media. Copyrights for components of this work owned by others than ACM must be honored. Abstracting with credit is permitted.
    	This is the author's version of the work.
    	The final authenticated version is available online at \href{https://doi.org/10.1145/3583678.3596888}{https://doi.org/10.1145/3583678.3596888} % will be available online
    	and has been published in the proceedings of the 
    	\confname.}
}
\begin{abstract}
Smart contracts have transformed decentralized finance by enabling programmable, trustless transactions. However, their widespread adoption and growing financial significance have attracted persistent and sophisticated threats, such as phishing campaigns and contract-level exploits. Traditional transaction-based threat detection methods often expose sensitive user data and interactions, raising privacy and security concerns. In response, static bytecode analysis has emerged as a proactive mitigation strategy, identifying malicious contracts before they execute harmful actions.

Building on this approach, we introduced \ph, the first machine-learning-based framework for detecting phishing activities in smart contracts via static bytecode and opcode analysis, achieving approximately 90\% detection accuracy. Nevertheless, two pressing challenges remain: (1) the increasing use of sophisticated bytecode obfuscation techniques designed to evade static analysis, and (2) the heterogeneity of blockchain environments requiring platform-agnostic solutions.

This paper presents a vision for \sys (\sysacronym), a robust, modular, and platform-agnostic framework for smart contract malware detection. Over the next 2.5 years, \sys will evolve in two stages: first, by tackling obfuscated Ethereum Virtual Machine (EVM) bytecode through graph neural network (GNN) analysis of control flow graphs (CFGs), leveraging GNNs’ ability to capture complex structural patterns beyond opcode sequences; and second, by generalizing detection capabilities to emerging runtimes such as WASM. \sys aims to enable proactive, scalable security for the future of decentralized ecosystems.
\end{abstract}


\section{Introduction}
Smart contracts form the core logic underpinning a wide range of decentralized applications (dApps). The Ethereum ecosystem, in particular, has experienced exponential growth—amassing billions of dollars in total value locked (TVL) and enabling innovative use cases such as lending pools, yield aggregators, and token swaps. However, the same programmable and trustless capabilities that drive these advancements also create opportunities for malicious actors. Recent estimates indicate that, in 2023 alone, Ethereum suffered over \$1.35 billion in financial losses due to security breaches, with phishing attacks and malicious smart contracts playing a significant role~\cite{defireport, chainabuse}.

Traditional approaches to detecting malicious contracts often rely on analyzing on-chain user interactions (e.g., transaction flows, approval patterns, or external calls). While effective under certain conditions, these methods require substantial transaction data to generate meaningful signals, thus delaying detection. Moreover, replaying or simulating transaction traces can inadvertently expose sensitive user data, creating additional security and privacy challenges. Given that blockchain transactions are irreversible, detection delays can result directly in monetary loss and erosion of user trust.

Static analysis of smart contract bytecode offers a proactive alternative: by detecting suspicious code structures at or before deployment, threats can be mitigated early. Motivated by this paradigm, we previously introduced \ph, a machine learning framework that analyzes smart contract bytecode to detect phishing attacks, achieving an average of 90\% detection accuracy. However, despite its promising results, \ph faces two limitations. First, emerging obfuscation techniques—including source-level manipulation~\cite{DBLP:journals/tse/ZhangYXDLWZ23}, control-flow graph transformations~\cite{DBLP:conf/apsec/YuZDXJ22}, and binary diversification~\cite{journals/cose/Cabrera-Arteaga24}—threaten the reliability of static pattern-based detection. Second, the increasing diversity of blockchain platforms (e.g., WASM-based smart contracts) demands solutions that are robust across heterogeneous runtimes.

In this paper, we present our vision for \sys (\sysacronym), a robust, platform-agnostic framework for malware detection in smart contracts. \sys will address obfuscation challenges by moving beyond opcode-level patterns, leveraging graph neural networks (GNNs) over control-flow graphs (CFGs) extracted from contract bytecode. GNNs offer the potential to learn high-level structural patterns in code execution flow, which may exhibit greater robustness against certain forms of obfuscation.

Our research plan unfolds over two phases. In the next six months, we aim to develop a robust detection method targeting obfuscated EVM bytecode, focusing on graph-based representations and machine learning models such as GAT~\cite{DBLP:conf/iclr/VelickovicCCRLB18}, GCN~\cite{DBLP:conf/iclr/KipfW17}, GIN~\cite{DBLP:conf/iclr/XuHLJ19}, TAG~\cite{DBLP:journals/corr/abs-1710-10370}, and GraphSAGE~\cite{DBLP:conf/nips/HamiltonYL17}. Over the subsequent two years, we plan to generalize \sys to support emerging smart contract runtimes, including WASM environments, building a modular and scalable detection pipeline. 

Through \sys, we aspire to proactively strengthen smart contract security across the evolving blockchain landscape.

\section{Background and Related Work}
The research community has developed various strategies to counter vulnerabilities in smart contracts. Dynamic or \emph{transaction-based} methods, such as Eth-PSD~\cite{ethpsd} and TxPhishScope~\cite{txphishscope}, monitor on-chain activity and external interactions to surface attacks in real time. While effective, these approaches require substantial user activity and may compromise privacy when replaying transactions.

Conversely, \emph{static analysis} techniques aim to detect threats before execution. Symbolic execution has been applied to uncover honeypots~\cite{honeypots} and hidden backdoors, but can be computationally intensive on complex contracts. Machine learning methods, including random forests, k-NN, and deep neural networks, have been used to classify attack types such as Ponzi schemes~\cite{DBLP:journals/jpscp/LiuPF23} or malicious opcode patterns~\cite{DBLP:conf/infocom/HuBX22}.

A tension persists between these approaches: dynamic methods benefit from live data but suffer from detection latency and privacy risks, while static methods offer proactive security but are vulnerable to code obfuscation. As DeFi ecosystems evolve, more robust and adaptive detection frameworks are urgently needed.

\section{\ph}
To address these challenges, we previously proposed \ph~\cite{DeRosaQBSF25}, a machine learning framework that detects phishing attacks through direct analysis of smart contract bytecode and opcode sequences. By focusing on code structure rather than transaction behavior, \ph enables proactive threat identification without requiring user interaction data.

\ph was evaluated on a curated dataset of 7,000 real-world phishing contracts, benchmarking 16 classification models, including opcode histograms, vision-based encodings, large language models, and vulnerability-specific detectors. Our approach achieved an average detection accuracy of approximately 90\%, demonstrating the promise of static, privacy-preserving malware detection at the bytecode level.

While effective, \ph faces two growing challenges: resilience against sophisticated obfuscation tactics and the need for broader applicability beyond the Ethereum Virtual Machine (EVM). These challenges motivate the development of \sys, a more general and robust detection framework.

\section{Existing Challenges: Obfuscation and Platform Heterogeneity}
\label{sec:limitations}

Although \ph proved effective in detecting phishing attacks on Ethereum, two major challenges limit static bytecode-based detection.

First, increasingly sophisticated obfuscation techniques, such as those introduced by BOSC~\cite{DBLP:conf/apsec/YuZDXJ22} and BiAn~\cite{DBLP:journals/tse/ZhangYXDLWZ23}, manipulate control structures, instruction flows, and data layouts to evade conventional static analyzers. These tactics reduce the reliability of opcode-pattern classifiers and shallow ML models.

Second, blockchain platforms are diversifying. Beyond the Ethereum Virtual Machine (EVM), ecosystems like Polkadot~\cite{polkadot}, NEAR~\cite{near}, the Internet Computer~\cite{dfinity} and EOS~\cite{eosio} rely on WebAssembly (WASM) as a runtime~\cite{immunebytes}, introducing heterogeneous instruction sets and execution models.

These dual challenges—advanced obfuscation and platform variability—necessitate a new class of resilient, adaptable, and cross-platform detection frameworks.

\section{Roadmap: From \ph to \sys}
\label{sec:roadmap}

To address these challenges, we propose \sys (\sysacronym), a next-generation malware detection framework for smart contracts. \sys will enhance robustness against obfuscation and scale across heterogeneous runtimes.

Our research unfolds in two phases:

\subsection{Phase 1: Robust Detection on Obfuscated EVM Bytecode}
In the next six months, we aim to detect obfuscated phishing contracts by moving beyond flat opcode sequences to control-flow graphs (CFGs). We will apply graph neural networks (GNNs)—including GCN, GAT, GIN, TAG, and GraphSAGE—over CFGs, exploring their potential to learn structural patterns that may be more resilient to superficial obfuscation.

We will also expand our existing 7,000-contract dataset by curating additional labeled samples from Etherscan~\cite{etherscan}, removing duplicates (e.g., minimal proxies~\cite{proxy}) to ensure diversity.

\subsection{Phase 2: Platform-Agnostic Detection}
Over the following two years, we will generalize \sys to support additional smart contract platforms beyond EVM, with a particular focus on WASM-based environments. This will involve adapting our analysis methods to different bytecode formats and execution models while maintaining consistent detection performance across platforms.

Ultimately, \sys aims to provide proactive, privacy-preserving security across decentralized ecosystems.

\section{Conclusion and Outlook}
We outlined our vision for \sys (\sysacronym), a bytecode-centric, platform-agnostic framework for smart contract malware detection. Building on \ph, \sys tackles two key challenges: resilience to obfuscation and adaptability across different execution environments.

Our two-phase roadmap first focuses on robust detection of obfuscated EVM bytecode using graph neural networks (GNNs) over control-flow graphs (CFGs), followed by expanding support to additional smart contract platforms such as WASM. Through \sys, we aim to enable proactive smart contract security and strengthen the protection of next-generation decentralized applications.

\bibliographystyle{plain}
\bibliography{biblio}

\begin{thebibliography}{10}

\bibitem{journals/cose/Cabrera-Arteaga24}
Javier Cabrera-Arteaga, Nicholas Fitzgerald, Martin Monperrus, and Benoit
  Baudry.
\newblock Wasm-mutate: Fast and effective binary diversification for
  webassembly.
\newblock {\em Computers \& Security}, 139:103731, 2024.

\bibitem{chainabuse}
ChainAbuse.
\newblock Chainabuse.
\newblock \href{https://www.chainabuse.com/}{https://www.chainabuse.com/}.

\bibitem{defireport}
De.Fi.
\newblock De.fi rekt report.
\newblock
  \href{https://de.fi/blog/de-fi-rekt-report-crypto-losses-reach-1-95b-in-2023}{https://de.fi/blog/de-fi-rekt-report-crypto-losses-reach-1-95b-in-2023}.

\bibitem{dfinity}
Dfinity.
\newblock The internet computer.
\newblock \href{https://developers.eos.io/}{https://developers.eos.io/}.

\bibitem{DBLP:journals/corr/abs-1710-10370}
Jian Du, Shanghang Zhang, Guanhang Wu, Jos{\'{e}} M.~F. Moura, and Soummya Kar.
\newblock Topology adaptive graph convolutional networks.
\newblock {\em CoRR}, abs/1710.10370, 2017.

\bibitem{eosio}
EOS.
\newblock The eos blockchain.
\newblock \href{https://developers.eos.io/}{https://developers.eos.io/}.

\bibitem{etherscan}
Etherscan.
\newblock Etherscan.io website.
\newblock \url{https://etherscan.io/}.
\newblock Accessed: 2024-11-25.

\bibitem{DBLP:conf/nips/HamiltonYL17}
William~L. Hamilton, Zhitao Ying, and Jure Leskovec.
\newblock Inductive representation learning on large graphs.
\newblock In Isabelle Guyon, Ulrike von Luxburg, Samy Bengio, Hanna~M. Wallach,
  Rob Fergus, S.~V.~N. Vishwanathan, and Roman Garnett, editors, {\em Advances
  in Neural Information Processing Systems 30: Annual Conference on Neural
  Information Processing Systems 2017, December 4-9, 2017, Long Beach, CA,
  {USA}}, pages 1024--1034, 2017.

\bibitem{txphishscope}
Bowen He, Yuan Chen, Zhuo Chen, Xiaohui Hu, Yufeng Hu, Lei Wu, Rui Chang, Haoyu
  Wang, and Yajin Zhou.
\newblock Txphishscope: Towards detecting and understanding transaction-based
  phishing on ethereum.
\newblock In {\em Proceedings of the 2023 ACM SIGSAC Conference on Computer and
  Communications Security}, CCS '23, page 120–134, New York, NY, USA, 2023.
  Association for Computing Machinery.

\bibitem{DBLP:conf/infocom/HuBX22}
Huiwen Hu, Qianlan Bai, and Yuedong Xu.
\newblock Scsguard: Deep scam detection for ethereum smart contracts.
\newblock In {\em {IEEE} {INFOCOM} 2022 - {IEEE} Conference on Computer
  Communications Workshops, {INFOCOM} 2022 - Workshops, New York, NY, USA, May
  2-5, 2022}, pages 1--6. {IEEE}, 2022.

\bibitem{immunebytes}
ImmuneBytes.
\newblock Revolutionizing blockchain: The impact of webassembly (wasm) on smart
  contracts.
\newblock
  \href{https://immunebytes.com/blog/revolutionizing-blockchain-the-impact-of-webassembly-wasm-on-smart-contracts/}{https://immunebytes.com/blog/revolutionizing-blockchain},.

\bibitem{ethpsd}
Arkan Hammoodi~Hasan Kabla, Mohammed Anbar, Selvakumar Manickam, and Shankar
  Karupayah.
\newblock Eth-psd: A machine learning-based phishing scam detection approach in
  ethereum.
\newblock {\em IEEE Access}, 10:118043--118057, 2022.

\bibitem{DBLP:conf/iclr/KipfW17}
Thomas~N. Kipf and Max Welling.
\newblock Semi-supervised classification with graph convolutional networks.
\newblock In {\em 5th International Conference on Learning Representations,
  {ICLR} 2017, Toulon, France, April 24-26, 2017, Conference Track
  Proceedings}. OpenReview.net, 2017.

\bibitem{DBLP:journals/jpscp/LiuPF23}
Derek Liu, Francesco Piccoli, and Victor Fang.
\newblock Machine learning approach to identify malicious smart contract
  opcodes: A preliminary study.
\newblock {\em JPS Conference Proceedings (Blockchain Kaigi 2023)}, 43:011002,
  2023.

\bibitem{near}
NEAR.
\newblock The near blockchain.
\newblock \href{https://near.org/}{https://near.org/}.

\bibitem{polkadot}
Polkadot.
\newblock The polkadot blockchain.
\newblock \href{https://polkadot.com/}{https://polkadot.com/}.

\bibitem{proxy}
Ethereum~Improvement Proposals.
\newblock Erc-1167: Minimal proxy contract.
\newblock \url{https://eips.ethereum.org/EIPS/eip-1167/}.
\newblock Accessed: 2024-11-25.

\bibitem{DeRosaQBSF25}
Pasquale~De Rosa, Simon Queyrut, Yerom{-}David Bromberg, Pascal Felber, and
  Valerio Schiavoni.
\newblock Phishinghook: Catching phishing ethereum smart contracts leveraging
  evm opcodes.
\newblock In {\em Proceedings of the 55th IEEE/IFIP International Conference on
  Dependable Systems and Networks (DSN)}, 2025.
\newblock Accepted for publication on 2025-03-19.

\bibitem{honeypots}
Christof~Ferreira Torres, Mathis Steichen, and Radu State.
\newblock The art of the scam: Demystifying honeypots in ethereum smart
  contracts.
\newblock In {\em 28th USENIX Security Symposium (USENIX Security 19)}, pages
  1591--1607, Santa Clara, CA, August 2019. USENIX Association.

\bibitem{DBLP:conf/iclr/VelickovicCCRLB18}
Petar Velickovic, Guillem Cucurull, Arantxa Casanova, Adriana Romero, Pietro
  Li{\`{o}}, and Yoshua Bengio.
\newblock Graph attention networks.
\newblock In {\em 6th International Conference on Learning Representations,
  {ICLR} 2018, Vancouver, BC, Canada, April 30 - May 3, 2018, Conference Track
  Proceedings}. OpenReview.net, 2018.

\bibitem{DBLP:conf/iclr/XuHLJ19}
Keyulu Xu, Weihua Hu, Jure Leskovec, and Stefanie Jegelka.
\newblock How powerful are graph neural networks?
\newblock In {\em 7th International Conference on Learning Representations,
  {ICLR} 2019, New Orleans, LA, USA, May 6-9, 2019}. OpenReview.net, 2019.

\bibitem{DBLP:conf/apsec/YuZDXJ22}
Qifan Yu, Pengcheng Zhang, Hai Dong, Yan Xiao, and Shunhui Ji.
\newblock Bytecode obfuscation for smart contracts.
\newblock In {\em 29th Asia-Pacific Software Engineering Conference, {APSEC}
  2022, Virtual Event, December 6-9, 2022}, pages 566--567. {IEEE}, 2022.

\bibitem{DBLP:journals/tse/ZhangYXDLWZ23}
Pengcheng Zhang, Qifan Yu, Yan Xiao, Hai Dong, Xiapu Luo, Xiao Wang, and Meng
  Zhang.
\newblock Bian: Smart contract source code obfuscation.
\newblock {\em {IEEE} Transactions on Software Engineering}, 49(9):4456--4476,
  2023.

\end{thebibliography}

\end{document}